\documentclass[12pt,preprint]{aastex}

\def\lya{Ly$\alpha$}

\def\kms{km s$^{-1}$}

\def\ergcm2s{\ifmmode {\rm\,erg\,cm^{-2}\,s^{-1}}\else
                ${\rm\,ergs\,cm^{-2}\,s^{-1}}$\fi}
\newcommand{\fluxunit}{\ergcm2s}

\begin{document}

\title{X-ray nondetection of the Ly$\alpha$ Emitters at $z$ $\sim$ 4.5}

\author{J. X. Wang\altaffilmark{1}, J. E. Rhoads\altaffilmark{2}, S. Malhotra\altaffilmark{2}, S. Dawson\altaffilmark{3}, D. Stern\altaffilmark{4}, A. Dey\altaffilmark{5}, T. M. Heckman\altaffilmark{1}, C. A. Norman\altaffilmark{1,2}, H. Spinrad\altaffilmark{3}}
\begin{abstract}

The \lya\ emitters found at $z \sim 4.5$ by the Large Area Lyman
Alpha (LALA) survey have high equivalent widths in the \lya\ line,
which can be produced by either narrow-lined active galactic nuclei (AGNs)
or by stellar populations with a very high proportion of young,
massive stars. To investigate the AGN scenario, 
we obtained two deep $Chandra$ exposures to study the
X-ray nature of the \lya\ emitters.
The 172 ks deep $Chandra$ image on the LALA Bo\"{o}tes field 
was presented in a previous paper (Malhotra et al. 2003),
and in this paper we present a new $Chandra$ deep exposure (174 ks)
on the LALA Cetus field, which doubled our sample of X-ray
imaged \lya\ sources, and imaged the brightest source among our
\lya\ emitters.
None of the 101 \lya\ sources covered by two $Chandra$ exposures
was detected individually in X-ray, with a $3 \sigma$ limiting
X-ray flux of $F_{\rm 0.5-10.0 keV} < 3.3 \times 10^{-16}$ \fluxunit\ 
for on-axis targets.
The sources remain undetectable in the stacked image, implying
a 3$\sigma$ limit to the average luminosity of $L_{\rm 2-8keV} < 2.8 \times 10^{42}$ ergs s$^{-1}$.
The resulting X-ray to \lya\ ratio is $>$ 21 times lower than the
ratios for known high redshift type-II quasars. 
Together with optical spectra obtained at Keck, we conclude that
no evidence of AGN activity was found among our \lya\ emitters
at $z \sim 4.5$.
\end{abstract}


\altaffiltext{1}{Department of Physics and Astronomy, Johns Hopkins University, 3400 N. Charles Street, Baltimore, MD 21218; jxw@pha.jhu.edu, heckman@pha.jhu.edu, norman@stsci.edu.}
\altaffiltext{2}{Space Telescope Science Institute, 3700 San Martin Drive, Baltimore, MD 21218; san@stsci.edu, rhoads@stsci.edu.} 
\altaffiltext{3}{Department of Astronomy, University of California at Berkeley, Mail Code 3411, Berkeley, CA 94720; sdawson@astron.berkeley.edu, spinrad@astro.berkeley.edu.}
\altaffiltext{4}{Jet Propulsion Laboratory, California Institute of Technology, Mail Stop 169-506, Pasadena, CA 91109; stern@zwolfkinder.jpl.nasa.gov.}
\altaffiltext{5}{National Optical Astronomy Observatory, 950 North Cherry Avenue, Tucson, AZ 85719; dey@noao.edu}
\section {Introduction}
More than three decades ago Partridge \& Peebles (1967) predicted
that galaxies undergoing their first throes of star formation 
should be strong emitters in the \lya\ line.
Recently, narrow-band surveys have been successful in finding \lya\
emitters at ever increasingly high redshifts.
These include our Large Area
Lyman Alpha (LALA) survey (e.g., Rhoads et al. 2000) and other recent
searches over smaller volumes (Cowie \& Hu 1998; Hu et al 1998, 2002, 2004;
Kudritzki et al 2000; Fynbo, Moller, \& Thomsen 2001; Pentericci et
al. 2000; Stiavelli et al. 2001; Ouchi et al. 2003; Fujita et
al. 2003, Shimasaku et al. 2003, Kodaira et al. 2003).  
The equivalent widths (EW) of
\lya\ emitters selected using narrow-band surveys tend to be large
(Malhotra \& Rhoads 2002, hereafter MR02; Kudritzki et al. 2000).
With the LALA survey we have found about 400 \lya\ candidates at 
$z \sim 4.5$, 101 of which are covered by our deep $Chandra$ images 
(Malhotra et al. 2003, hereafter M03; and this letter).
We have obtained optical spectra for a subset of the
\lya\ candidates, resulting
in 47 spectroscopical confirmations to date and a spectroscopical
success rate of $>$ 72\% (Dawson et al. 2004, in prep).
The median equivalent width of the \lya\ line
is greater than 240\AA\ in the LALA $z \sim 4.5$ sample (MR02), and
comparable at z $\sim$ 5.7 (Rhoads \& Malhotra 2001, Rhoads et al. 2003).
Hu et al. (2004) have found a somewhat smaller fraction (25\% with
EW $>$ 240\AA) at z $\sim$ 5.7 based on a sample that is spectroscopically
confirmed but much smaller than the MR02 sample.
Normal stellar populations can produce \lya\ emission with equivalent width
240\AA\ or less (Charlot \& Fall 1993), unless they have 
a top-heavy initial mass function (IMF),
zero (or very low) metallicity, and/or extreme youth (age $< 10^7$
years).
By comparison, stellar populations older than 10$^7$ years and with a Salpeter IMF = 2.35 can only produce \lya\ emission with equivalent width of $\sim 100$ \AA\ (see Fig.2 of MR02).
The high equivalent widths could also be explained if narrow-lined 
(type II) active
galactic nuclei (AGNs) were present in our \lya\ emitter sample.
The broad-lined (type I) AGNs are ruled out because we see no evidence
of broad emission lines from either narrow-band imaging or spectroscopy.

A large population of high redshift \lya\ emitting AGNs would have
interesting implications both for the pace of black hole formation
and for cosmic background radiation from the
gamma ray to the far infrared.  Of particular interest is the
possibility of a large population of type II quasars, i.e., 
high luminosity AGNs ($L_{\rm X} >$ 10$^{44}$ ergs s$^{-1}$)
whose broad line regions and soft X-rays are greatly attenuated by
large column densities of gas and dust.  
Deep X-ray surveys have uncovered the first X-ray selected type~II quasars
(Norman et al 2002, Stern et al. 2002, Dawson et al. 2003, Della Ceca et al. 2003), and
more candidates (i.e., Fiore et al. 2003, Crawford et al. 2000, Mainieri et al.
2002).
Inspired by these examples, and the fact that the \lya\ line luminosities
of the LALA z $\sim$ 4.5 candidates are comparable to these of the high 
redshift type~II quasars, we set out to image the LALA survey fields with 
$Chandra$ Advanced CCD Imaging Spectrometer (ACIS) to explore the possibility 
that some of the \lya\ emitters might be type II quasars. 

Two deep $Chandra$ ACIS images were obtained to check for the presence
of type II quasars among our \lya\ emitters.
The results of the 172 ks $Chandra$ deep image on the LALA Bo\"{o}tes field
was provided by M03. In this paper,
we present a new 174 ks $Chandra$ deep image on LALA Cetus field,
which doubled our sample to 101 X-ray imaged \lya\ emitters.
The new $Chandra$ image was also designed to cover the ``boomer"
of our \lya\ emitters: LALA J020438.95-051116.2, 
which is the brightest confirmed \lya\ emitter selected by
by the LALA survey, with line flux of $9.3 \times 10^{-17} \fluxunit$ and a 
rest frame equivalent width of $650$~\AA.

\section{Optical and $Chandra$ imaging}
The LALA survey comprises two primary fields, located in Bo\"{o}tes
(J142557+3532) and in Cetus (J020520--0455).
Each field is $36\arcmin \times 36\arcmin$  in size,
corresponding to a single field of the 8192$\times$8192 pixel Mosaic CCD
cameras at the National Optical Astronomy Observatory's 4 meter telescopes.
For each field, 5 partially overlapping narrow band filters covering 
$4.37 < z < 4.57$ for \lya\ were utilized to obtain deep narrowband
images. These filters have full
width at half maximum (FWHM) $\approx 80$\AA\ and are spaced at
$\approx 40$\AA\ intervals (central wavelengths $\lambda_c \approx
6559$, $6611$, $6650$, $6692$, and $6730$\AA).  
For the Bo\"{o}tes field, 3 more narrowband images covering 
$z \sim 5.7$ (Rhoads et al. 2003) and $z \sim 6.5$ (Rhoads et al. 2004) 
are also available.
Imaging data reduction followed the methods described in Rhoads et al
(2000), and \lya\
candidates were selected using criteria described by Rhoads \& Malhotra (2001).
This resulted in a total of $\sim$ 400 good candidates at $z \sim 4.5$,
which will be presented in a future paper (Rhoads et al. 2004,
in preparation).

Two deep $Chandra$ ACIS images were obtained for the LALA Bo\"{o}tes
and Cetus fields.
The $Chandra$ observations were designed to maximize the number of 
large equivalent width sources within the ACIS-I field of view.
The 172 ks (net exposure) $Chandra$ exposure of the  LALA Bo\"{o}tes field
was taken in 2002, and details of the X-ray data analyses can be found in 
M03 and Wang et al. (2004).
The new 174 ks (net exposure) $Chandra$ ACIS exposure on the LALA Cetus field
was obtained in very faint (VFAINT) mode on UT 2003 June 13-15.
Data reduction was done with the package CIAO 2.2.1 (see
http://asc.harvard.edu/ciao), following the procedures described in M03.
The average offset between the X-ray and optical images was obtained by
comparing X-ray source positions with optical counterparts
(whenever found). The derived 0.3$\arcsec$ offset has been 
applied for all subsequent analyses.
No obvious rotation and plate-scale effects were discovered.
We ran WAVDETECT (Freeman et al. 2002) on the soft (0.5 -- 2.0 keV), 
hard (2.0 -- 7.0 keV), and total band (0.5 -- 7.0 keV) X-ray images. 
The detailed X-ray data reduction and the detected X-ray sources will be
presented in a future paper (Wang et al. 2004, in preparation).

\section{X-ray imaging Results}
\subsection{Non-detection of individual sources}
\begin{figure}
\plotone{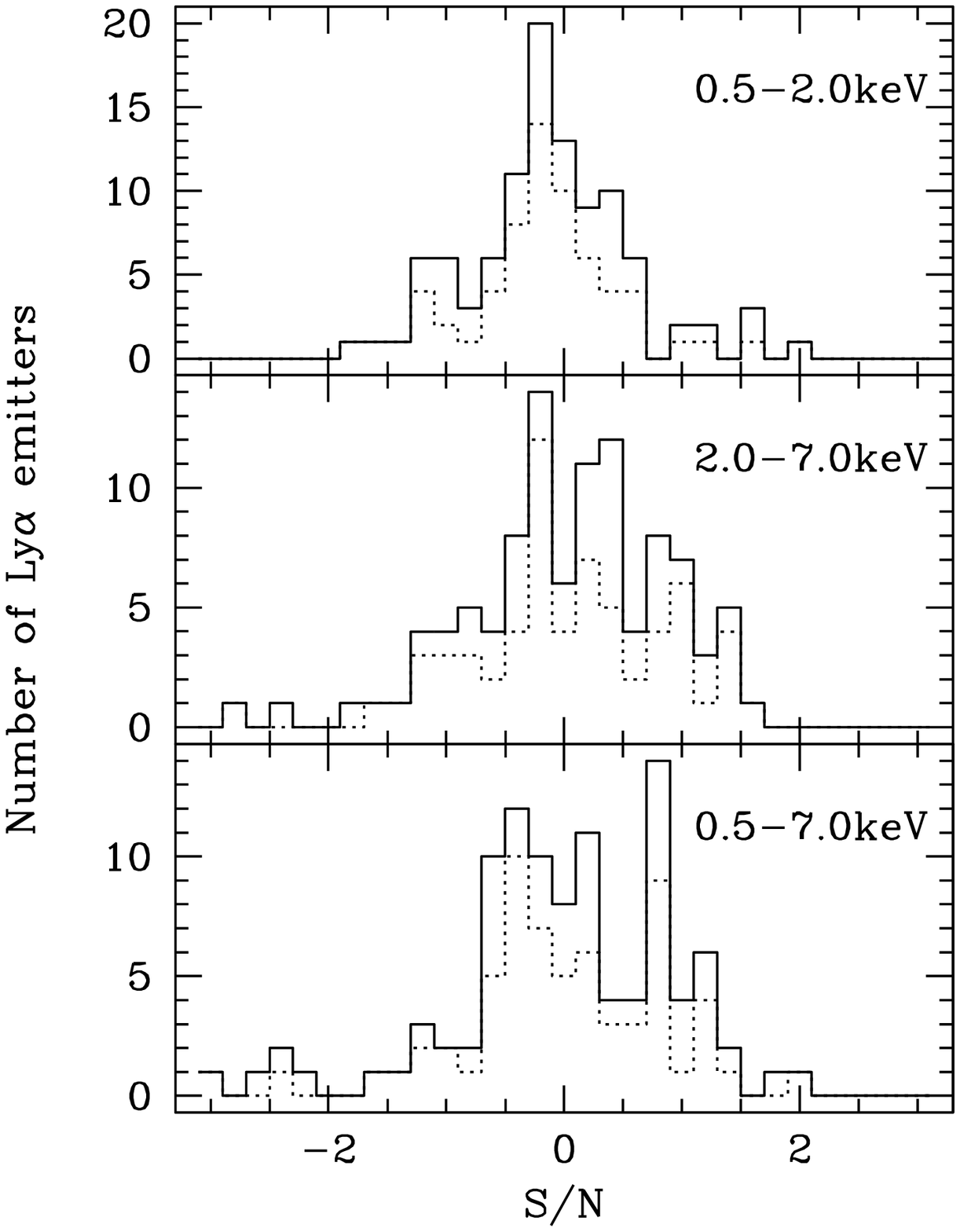}
\caption{
Histogram distributions of the X-ray signal-to-noise ratios (S/N)
for \lya\ sources in the 0.5 -- 2.0 keV, 2.0 -- 7.0 keV and 0.5 -- 7.0 keV 
band. The total
sample of 101 \lya\ sources is shown as the solid line.  The sub-sample of
64 \lya\ sources with rest frame EWs $>$ 240 \AA\ is overplotted as the 
dotted line. None of the \lya\ sources has an X-ray net count above
2 $\sigma$ level in any band.
}
\label{hist}
\end{figure}

A total of 101 \lya\ sources were imaged by the two $Chandra$ 
exposures, 49 in the Bo\"{o}tes field, and 52 in the Cetus field.
None were detected.
To perform X-ray photometry analyses, for each \lya\ source
we defined a circular source region 
centered at the source position on the X-ray images, with radius R$_s$ set
to the 95\% encircled-energy radius of {\it Chandra\/} ACIS PSF at the
position. 
The source regions defined above were used to extract source
photons for the \lya\ sources, and the backgrounds were extracted
from an annulus with $1.2 R_s < R < 2.4 R_s$ after masking out nearby 
X-ray sources. The net count of each source was derived after accounting
for differences of effective exposure time between source regions
and background regions (mainly due to CCD edge effects and bad columns).
We found that none of the \lya\ sources has an X-ray net count above
2 $\sigma$ level in either 0.5 -- 2.0 keV, 2.0 -- 7.0 keV or 0.5 -- 7.0 keV band. In Fig. \ref{hist} 
we present the histogram distributions of the signal-to-noise ratios (S/N)
of the X-ray net counts from \lya\ sources.
 
We conclude that none of the \lya\ sources are detected by the
X-ray observations, allowing us to place upper limits on their
X-ray fluxes.  For the \lya\ source nearest
to the axis of X-ray observation (with an off-axis angle of
1.7$\arcmin$), there are no photons
within R$_s$ in the 0.5 -- 7.0 keV band.  
The 3$\sigma$ level upper limit of X-ray count is
6.61 (Gehrels 1986). 
Assuming a powerlaw spectrum with photon index of 2.0, this corresponds
to a 3$\sigma$ upper limit of X-ray flux $F_{\rm 0.5 - 10.0 keV} <
3.3 \times 10^{-16}$ \fluxunit.
Note that to make our fluxes in this paper directly comparable to those in 
other literatures,
we convert the 0.5 -- 7.0 keV band X-ray counts to $F_{\rm 0.5 - 10.0 keV}$.
Other sources have higher upper
limits due to the lower effective areas and larger PSF sizes.
The ``boomer", LALA J020438.95-051116.2, is located at off-axis angle
of 6.11\arcmin\ in the $Chandra$ image. 
We found no evidence of X-ray emission from the ``boomer": X-ray
photometry yields a net count of 1.0 in the 0.5 -- 7.0 keV band,
with 3$\sigma$ upper limit of 12.2.
The corresponding 3$\sigma$ upper limit to the X-ray flux is
$F_{\rm 0.5 - 10.0 keV} = 6.6 \times 10^{-16}$ \fluxunit.

\subsection{Stacking analysis}
\begin{figure}
\plotone{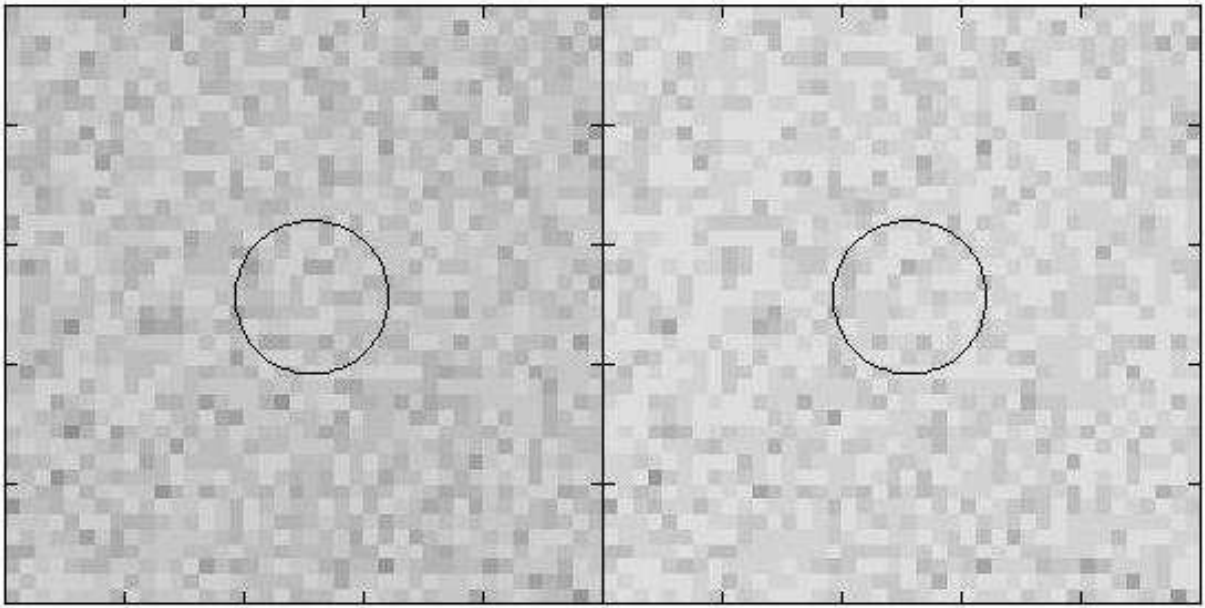}
\caption{
Stacked 0.5 -- 7.0 keV $Chandra$ images of \lya\ sources.
Left: all \lya\ emitters with off-axis angle $<$ 8\arcmin;
Right: \lya\ emitters with off-axis angle $<$ 8\arcmin\ and rest frame
equivalent widths EW $>$ 240\AA.
The effective exposure time of the stacked images
is 11.2 Ms (left panel). The images are $\sim$ 20\arcsec $\times$ 20\arcsec\
in size, and the circles are centered on the stacking positions with a radius of
2.5\arcsec.
}
\label{stack}
\end{figure}

The symmetric histograms in Fig. \ref{hist} suggest that X-ray emission
will not be detectable in summed images of either the total sample of
\lya\ sources, or of the sub-sample of 64 \lya\ sources with rest frame
EWs $>$ 240\AA. 
We performed a stacking analysis to calculate
the upper limit of the average X-ray flux for the \lya\ sources.
We note that the X-ray fluxes of \lya\ sources located at 
larger off-axis angles ($>$ 8\arcmin) are less constrained,
because of their larger PSF sizes and lower effective exposures.
Thus we only stacked the $Chandra$ images for the 70 \lya\ sources
(69\% of the whole sample) with off-axis angles $<$ 8\arcmin, 
providing an effective exposure time of $\sim$ 10.6 Ms.
The stacked 0.5 -- 7.0 keV image is presented in Fig. \ref{stack}.
The second panel of Fig.~2 presents the
stacked image for the subset of these sources with larger \lya\ equivalent
widths (EW $>$ 240 \AA\ in the rest frame).  No ~cumulative~ X-ray emission
is detected in either stacked image.
For each of the 70 sources, X-ray photons within
80\% encircled-energy radius of {\it Chandra\/} ACIS PSF were
extracted and summed up. After subtracting the local background
and applying aperture corrections, we obtained a net count of
-12 in the 0.5 -- 2.0 keV band, and -20 in the 0.5 -- 7.0 keV band, 
with 3$\sigma$ upper limits of 29 and 52 respectively.
The corresponding 3$\sigma$ upper limit on the average X-ray flux for
the $z \sim$ 5 \lya\ source population is 
$F_{\rm 0.5 - 2.0 keV} < 1.3 \times 10^{-17}$ \fluxunit,
which is 67\% of the limit in M03\footnote{
Using the 95\% encircled energy radius for the ACIS
PSF provides a somewhat degraded upper limit to the average X-ray flux.
The method for calculating the average X-ray flux upper limit described in
$\S$3.2 of M03 is somewhat inaccurate.  However, the limit presented there
is very close to the correct value determined using the method provided here.
},
and  $F_{\rm 0.5 - 10.0 keV} < 4.2 \times 10^{-17}$ \fluxunit.
Note that including sources with off-axis angles $>$ 8\arcmin\ will
increase the summed effective exposure time to $\sim$ 14 Ms;
however, it does not provide an improved upper limit on the average 
X-ray flux.

\section {Discussion and Conclusions}

We imaged 101 \lya\ emitters at z $\sim$ 4.5 with two
170 ks deep $Chandra$ ACIS exposures. None were clearly detected
at X-ray energies. 
In Fig. 3, we present the 1$\sigma$ upper limits
of the X-ray fluxes against their \lya\ line fluxes, which in most cases are 
the only well measured properties of the \lya\ sources.
For comparison, three high redshift type II quasars,
CDF-S202 ($z$ = 3.7, Norman et al. 2002), CXO52 ($z$ = 3.288, Stern et al. 2002) and HDFX28 ($z$ = 2.011, Dawson et al. 2003) are
also plotted in the figure.
We found that after scaling the X-ray fluxes with \lya\ line fluxes, all the 
\lya\ emitters at $z \sim 4.5$ are fainter in X-ray than HDFX28, 
$>$ 96\% of them are fainter than  CDF-S202,
and $>$ 50\% are fainter than CXO52. 
This indicates that $<$ 4\% of our \lya\ emitters at $z \sim 4.5$ can be 
type II quasars like CDF-S202, and $<$ 50\% of them can be like CXO52.
However, that average X-ray to \lya\ ratio (1$\sigma$ upper limit) is
 $>$ 21 times fainter than that of CXO52, which means actually $<$ 4.8\% of 
the \lya\ emitters could resemble CXO52.

Could some of our \lya\ emitters still be Seyfert 2 galaxies (i.e., lower 
luminosity type II AGNs)? 
Kriss (1984, 1985) presented samples of low redshift ($z<1$) Seyferts
and quasars with good measurements of \lya\ line and X-ray luminosities.
The AGNs in Kriss's samples have X-ray luminosities 
spreading from L$_X$ $\sim$ 10$^{42}$ erg s$^{-1}$ to 10$^{47}$ erg s$^{-1}$,
but no evidence of a correlation between their X-ray to \lya\ ratios and
X-ray luminosities was found.
This indicates that the X-ray to \lya\ ratios of low luminosity AGNs (i.e.,
Seyferts) are similar to these of luminous ones (i.e., quasars), thus the average X-ray to \lya\ ratio (upper limit) of our
\lya\ emitters is not only significantly fainter than luminous quasars,
but also lower luminosity AGNs.
The average X-ray to \lya\ ratio of Kriss's samples is also plotted in Fig. 3,
which is pretty close to these of the three type II quasars, and 8.5
times higher than the average ratio (upper limit) of our \lya\ emitters at 
$z \sim 4.5$. 

No evidence of AGN activity among our \lya\ emitters was found in
optical spectra either.
At $z \sim 4.5$, we have 18 confirmed 
\lya\ lines with Keck/LRIS, and 29 with Keck/DEIMOS (Dawson et al. 2004). 
The \lya\ lines are found to be
narrow ($<$ 500 \kms) in all the spectroscopically confirmed \lya\ emitters
(e.g., Rhoads et al. 2000, Rhoads et al. 2003, Dawson et al. 2004).
This is in fact narrower than the typical 
physical line widths of even type II quasars (see Rhoads et al 2003).
None of the spectra shows high excitation emission lines either (e.g., NV, CIV, HeII and CIII]),
even in the stacked spectrum (Dawson et al. 2004).
In Fig.4, we presented the Keck LRIS spectrum of our ``boomer", 
LALA J020438.95-051116.2, which has been confirmed to be normal
star forming galaxy at z=4.457. We can clearly see a narrow
(with a full width half maximum of $<$ 400 \kms) and asymmetric
\lya\ line, but no other lines indicating AGN activity.

We conclude that no evidence of AGN activity among our \lya\ emitters
at z $\sim$ 4.5 was found, from either deep $Chandra$ images, or
optical spectra.
The $3\sigma$ upper limit of 0.5 -- 2.0 keV flux on the average LALA sources
corresponds to an X-ray luminosity of $2.8 \times 10^{42}$ ergs s$^{-1}$ at 
$z \sim 4.5$ (for either 0.5 -- 2.0 keV rest frame or 2.0 -- 8.0 keV rest
frame bandpass, H0=65 km s$^{-1}$ Mpc$^{-1}$, $\Omega_m$=1/3,
$\Omega_\lambda$=2/3).  
The corresponding $3\sigma$ upper limits of L$_{\rm 2-8keV}$ for on-axis 
individual \lya\ emitters and the ``boomer" are 3.3 $\times 10^{43}$ and 6.6 $\times 10^{43}$ 
ergs s$^{-1}$ respectively.
Our analysis indicates that $<$ 4.8\% of our \lya\ emitters could
be possible AGNs based on their average X-ray to \lya\ ratio.
Steidel et al. (2002) found that 3\% of their Lyman break galaxies
at $z \sim 3$ show clear evidence for AGN activity.
It's interesting to note that the upper limit of AGN fraction of our
\lya\ emitters is well consistent with that of Lyman break galaxies.

\acknowledgements 
This work was supported by the CXC grant GO2-3152x and GO3-4148X, and
has benefited from images provided by the NOAO Deep Wide-Field
Survey (NDWFS; Jannuzi \& Dey 1999, Brown et al. 2003),
which is supported by the National
Optical Astronomy Observatory (NOAO).  NOAO is operated by AURA,
Inc., under a cooperative agreement with the National Science
Foundation. We would like to thank B. T. Jannuzi for his help on
narrow band imaging observations, and L. Kewley for
checking the \lya\ flux of CDF-S 202 in the observed frame.
The work of SD was supported by IGPP--LLNL University
Collaborative Research Program grant \#03--AP--015.
The work of DS was carried out at Jet Propulsion Laboratory, California Institute of
Technology, under a contract with NASA.
This work includes observations
made at the W.M. Keck Observatory, which is operated as a scientific
partnership among the California Institute of Technology, the University
of California and the National Aeronautics and Space Administration.

\clearpage

\newpage

\begin{figure}
\plotone{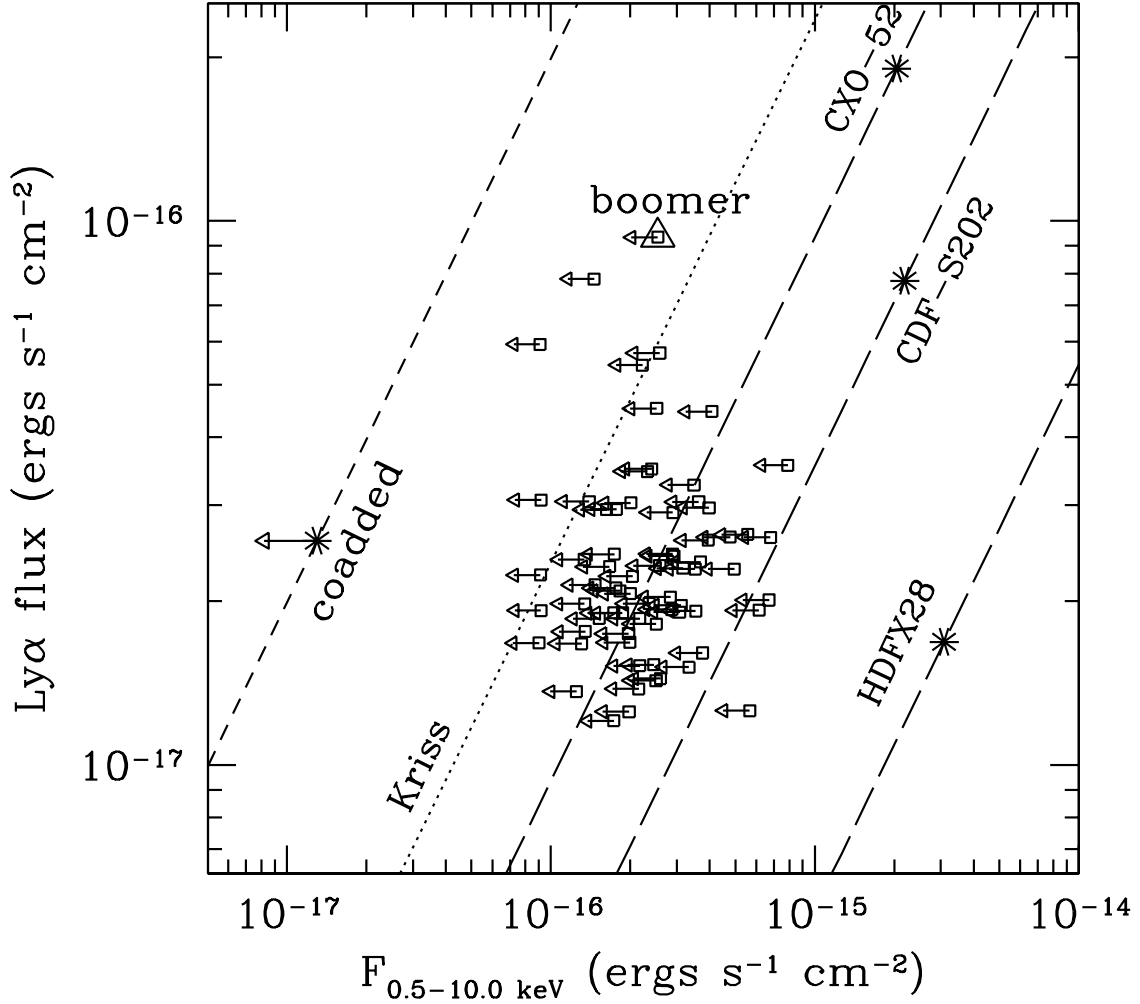}
\caption{
\lya\ fluxes vs 1$\sigma$ upper limits of the X-ray fluxes 
for our \lya\ sources at z $\sim$ 4.5, compared with three known high redshift
type II quasars. The ``boomer" of our \lya\ sources is plotted by
a big triangle. The average \lya\ flux is plotted against
the 1$\sigma$ upper limit of the average X-ray flux.
Lines indicate constant X-ray to \lya\ flux ratios.
The average X-ray to \lya\ ratio for the lower luminosity, low redshift 
($z<1$) Seyferts and quasars samples (Kriss 1984, 1985) is also plotted.
}
\end{figure}

\begin{figure}
\plotone{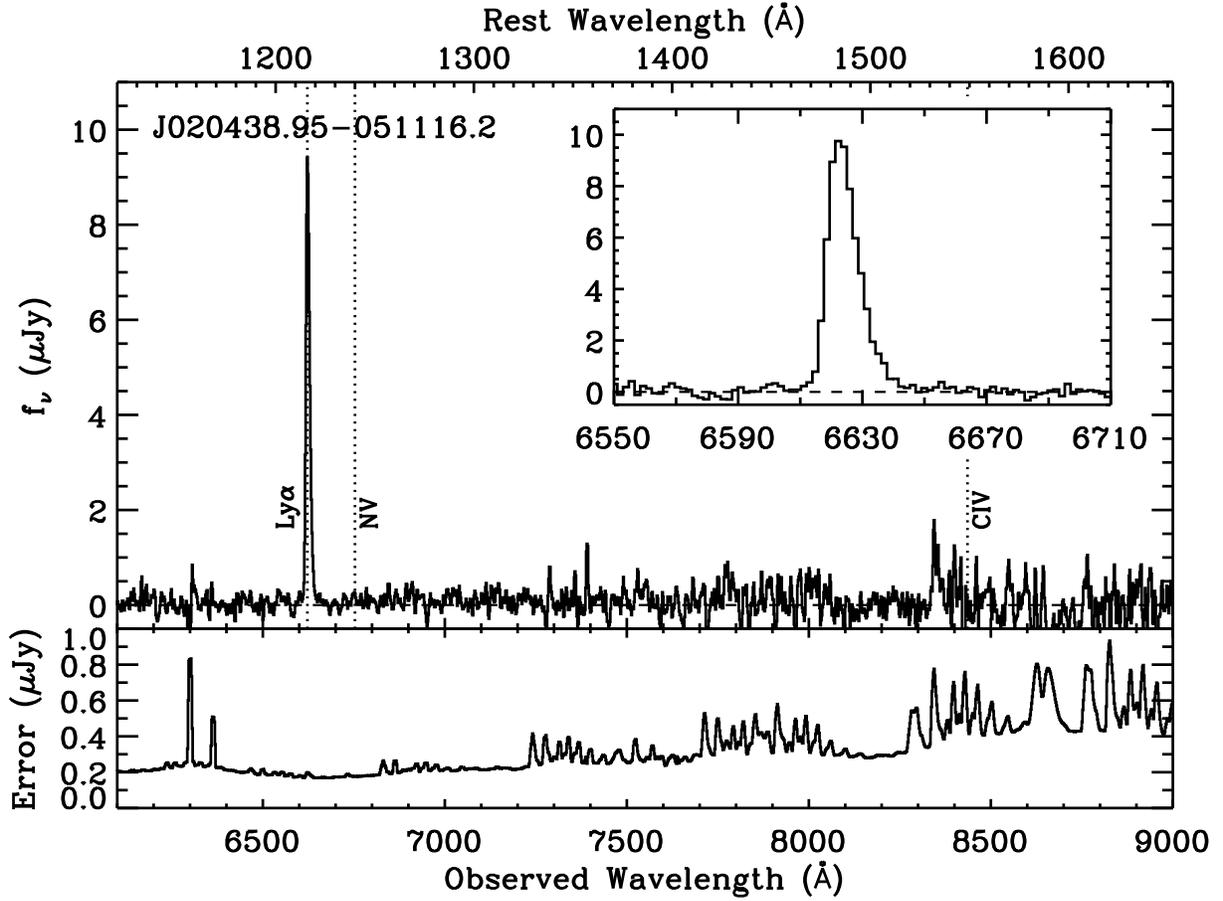}
\caption{
Keck LRIS Spectrum of the ``boomer" of our \lya\ emitters at z $\sim$ 4.5.
The source is confirmed to have a spectroscopic redshift of 4.457.
The photon counting errors are plotted in the bottom, and the insert shows 
the profile of the narrow and asymmetric \lya\ line.
}
\end{figure}
\end{document}